\documentclass[prd,showpacs,preprintnumbers,amsmath,amssymb]{revtex4}

\begin{document}

\title{Thermofield Dynamics of Time-Dependent Boson and Fermion Systems}

\author{Sang Pyo Kim}\email{sangkim@kunsan.ac.kr}

\affiliation{Department of Physics, Kunsan National University, Kunsan
573-701, Korea}

\affiliation{Asia Pacific Center for Theoretical
Physics, Pohang 790-784, Korea}

\author{F. C. Khanna}\email{khanna@phys.ualberta.ca}

\affiliation{Theoretical Physics Institute, Department of
Physics, University of Alberta, Edmonton, Alberta, Canada T6G
2J1}

\affiliation{TRIUMF, 4004 Wesbrook Mall, Vancouver, British Columbia,
Canada, V6T 2A3}

\date{\today}

\begin{abstract}
We formulate the thermofield dynamics for time-dependent
systems by combining the Liouville-von Neumann equation,
its invariant operators, and the basic notions of thermofield dynamics.
The new formulation is applied to time-dependent bosons and fermions by using  
the time-dependent annihilation and
creation operators that satisfy the Liouville-von Neumann equation.  
It is shown that the thermal state is the time- and
temperature-dependent vacuum state and a general formula is derived to
calculate the thermal expectation value of operators.
 
\end{abstract}

\maketitle

\section{Introduction}

The thermofield dynamics (TFD)  introduced by Takahashi and
Umezawa three decades ago is a canonical formalism for finite
temperature theory to describe quantum systems in thermal
equilibrium \cite{takahashi,umezawa1}. The great merit of TFD is
that it preserves many useful properties of the zero-temperature
field theory. The central concept in TFD is the thermal state, a
pure state, in some extended Hilbert space, which corresponds to
the thermal equilibrium, a mixed state, in the original Hilbert
space. The three essential ingredients of TFD are $(i)$ the tilde
conjugation rule, $(ii)$ the Heisenberg equation and $(iii)$ the
thermal state conditions \cite{umezawa2}. In a simple physical
terminology, the TFD of a quantum system doubles the degrees of
freedom by introducing a fictitious Hamiltonian
without any interaction with the system and uses an extended
Hilbert space of the direct product of the Hilbert spaces of the
system plus the fictitious system. In the oscillator
representation there is a temperature-dependent Bogoliubov
transformation between the annihilation and creation operators of
the total system and those of temperature-dependent ones. Then
the thermal state  is a
two-mode squeezed vacuum state (temperature-dependent vacuum state)
of the extended Hilbert space, which in turn is annihilated by the
temperature-dependent annihilation operators. There have been
numerous diverse applications of TFD to systems in condensed matter, 
nuclear physics, particle physics, quantum optics and cosmology 
in thermal equilibrium (for review and references, see \cite{umezawa2} 
and \cite{henning}).

On the other hand, there are also many systems exhibiting
nonequilibrium characteristics. An open system
interacting with a reservoir or a time-dependent system is such a
nonequilibrium system. As the time-translational invariance 
of such a time-dependent system is
broken, there is an obstacle in applying the Matsubara's
imaginary-time method \cite{matsubara}. Neither does the
conventional wisdom work well using the basis of the
time-dependent energy eigenstates in evaluating thermal quantities
since the density operator is not given by $ e^{- \beta H(t)}$.
However, the closed-time path method by Schwinger and Keldysh is
well-known and is widely used for such nonequilibrium 
systems \cite{schwinger}.
Another canonical theory based on the (functional)
Schr\"{o}dinger equation has been employed to study nonequilibrium
evolution of time-dependent systems \cite{jackiw}. Keeping the
three ingredients of TFD, requiring the Hamiltonian to be tildian, and
using the Heisenberg picture, Umezawa {\it et al} attempted to
extend the TFD to such nonequilibrium systems \cite{arimitsu}.

In this paper we show that when $(ii)$ the Heisenberg equation,
one of the ingredients of TFD, is replaced by the Liouville-von
Neumann (LvN) equation, the TFD has a direct generalization to
time-dependent boson and fermion systems, in particular, time-dependent
oscillators whose mass and frequency may change
explicitly in time. The invariant operators satisfying the
LvN equation not only lead to the Hilbert (Fock)
space of exact quantum states \cite{lewis} but also provide the
correct density operator \cite{kim1}. Thus we are able to extend
the TFD to time-dependent bosons (fermions) first by using the
time-dependent annihilation and creation operators for the
bosons (fermions) and the fictitious bosons (fermions), all linear invariant
operators \cite{kim1,lin inv,kim2,kim-khanna}, and then by finding the time-
and temperature-dependent annihilation and creation operators
through temperature-dependent Bogoliubov transformation of TFD.
Then the thermal state is a two-mode squeezed state of the
time-dependent vacuum state for the bosons (fermions) plus the fictitious
bosons (fermions). We find a general formula for evaluating the thermal
expectation values of operators and finally discuss the
distribution which evolves from an initial boson distribution.

The organization of this paper is as follows: In Sec. II, we
briefly review the TFD for boson and fermion systems. In Sec. III, we
introduce the time-dependent annihilation and creation operators
for a time-dependent boson. In Sec. IV, we extend the TFD to
time-dependent boson and discuss the physical
implication of the TFD for a boson which evolves from an
initial thermal state to a final one through a time-dependent
interaction. In Sec. V, TFD is extended to time-dependent fermions.
Additional comments and conclusion are given in Sec. VI.

\section{TFD for Boson and Fermion Systems in Equilibrium}

We briefly review the TFD for static (time-independent)
bosons and fermions in a way that can be readily applied to 
time-dependent ones in the following sections. The static
boson (fermion) has the Hamiltonian of the form
\begin{equation}
H = \hbar \omega a^{\dagger} a,
\end{equation}
where the standard annihilation and creation operators satisfy the
commutator (anticommutator)
\begin{equation}
[ a, a^{\dagger}]_{\pm} = 1.
\end{equation}
Quantum statistics of the boson (fermion) in thermal equilibrium is
described by the density operator
\begin{equation}
\rho = \frac{1}{Z} e^{- \beta H}, \quad Z = {\rm Tr} [e^{- \beta
H}],
\end{equation}
where $\beta$ is the inverse temperature. For instance, the
thermal expectation value of an operator $A$ is given by
\begin{eqnarray}
\langle A \rangle_{\rm T} = {\rm Tr} [\rho A] = \frac{1}{Z} \sum_n
e^{ - \beta \hbar \omega n} \langle n \vert A \vert n
\rangle,
\end{eqnarray}
with $\vert n \rangle$ being number states. The thermal
equilibrium is a mixed state with the probability $p_n = e^{-
\beta \hbar n }$ for each projector $\vert n \rangle
\langle n \vert$.

The idea of TFD is to double the system by adding a fictitious
system and extend the thermal equilibrium in the system's Hilbert
space to a thermal state, a pure state, in the extended Hilbert
space of the total system. For that purpose,
let us introduce a fictitious boson (fermion) using the tilde conjugation rule,
with the Hamiltonian,
\begin{equation}
\tilde{H} = \hbar \omega \tilde{a}^{\dagger} \tilde{a},
\end{equation}
with the number state $\vert \tilde{n} \rangle$ and the commutator (anticommutator)
\begin{equation}
[ \tilde{a}, \tilde{a}^{\dagger}]_{\pm} = 1.
\end{equation}
As $\{ a, a^{\dagger} \}$ and $\{\tilde{a}, \tilde{a}^{\dagger}
\}$ describe two independent systems, $H$ and $\tilde{H}$,
respectively, they commute (anticommute) with each other
\begin{equation}
[a, \tilde{a}]_{\pm} = [a, \tilde{a}^{\dagger}]_{\pm} = [a^{\dagger},
\tilde{a}]_{\pm} = [ a^{\dagger}, \tilde{a}^{\dagger}]_{\pm} = 0.
\end{equation}
The total Hamiltonian
\begin{equation}
\hat{H} := H - \tilde{H} = \hbar \omega (a^{\dagger} a -
\tilde{a}^{\dagger} \tilde{a})
\end{equation}
now carries an extended Hilbert space $\hat{\cal H} = {\cal H}
\otimes {\tilde{\cal H}}$. Each state of the extended Hilbert
space consists of a product of number states
\begin{equation}
\vert n, \tilde{m} \rangle = \vert n \rangle \otimes \vert
\tilde{m} \rangle.
\end{equation}

In TFD the thermal equilibrium corresponds to the thermal state in
the extended Hilbert space, described by a  pure state
\begin{eqnarray}
\vert 0(\beta) \rangle &=& \frac{1}{Z^{1/2}} \sum_n e^{- \beta
\hbar \omega n/2} \frac{1}{n!} a^{\dagger n} \tilde{a}^{\dagger n}
\vert 0, \tilde{0} \rangle \nonumber\\ &=& (1 \mp e^{- \beta
\omega})^{1/2} e^{- (\beta \hbar \omega/2) a^{\dagger}
\tilde{a}^{\dagger}} \vert 0 \rangle, \label{th vac}
\end{eqnarray}
where the upper (lower) sign is for bosons (fermions) and  
$ \vert 0 \rangle = \vert 0, \tilde{0} \rangle$ in the second line. 
In fact, the thermal state is a two-mode squeezed
vacuum state obtained by applying a Bogoliubov transformation to the vacuum state
\begin{equation}
\vert 0(\beta) \rangle = e^{- i G} \vert 0 \rangle, \label{sq st}
\end{equation}
where
\begin{equation}
G = - i \theta (\beta) (\tilde{a} a - a^{\dagger}
\tilde{a}^{\dagger}).
\end{equation}
Here $\beta (\theta)$ is a temperature-dependent parameter
determined by
\begin{eqnarray}
\cosh \theta (\beta) &=& (1 - e^{- \beta \hbar \omega} )^{-1/2},
\nonumber\\ \sinh \theta (\beta) &=& e^{- \beta \hbar \omega/2} (1 -
e^{- \beta \hbar \omega})^{-1/2}, \label{cosh}
\end{eqnarray}
for bosons and
\begin{eqnarray}
\cos \theta (\beta) &=& (1 + e^{- \beta \hbar \omega} )^{-1/2},
\nonumber\\ \sin \theta (\beta) &=& e^{- \beta \hbar \omega/2} (1 +
e^{- \beta \hbar \omega})^{-1/2}, \label{cos}
\end{eqnarray}
for fermions.

The two-mode squeeze operator introduces the temperature-dependent
annihilation and creation operators through the Bogoliubov
transformations
\begin{eqnarray}
a(\beta) &=& e^{- i G} a e^{i G} = \cosh \theta (\beta) a - \sinh
\theta (\beta) \tilde{a}^{\dagger}, \nonumber\\ \tilde{a}(\beta)
&=& e^{- i G} \tilde{a} e^{i G} = \cosh \theta (\beta) \tilde{a} -
\sinh \theta (\beta) a^{\dagger},
\end{eqnarray}
for bosons and
\begin{eqnarray}
a(\beta) &=& e^{- i G} a e^{i G} = \cos \theta (\beta) a - \sin
\theta (\beta) \tilde{a}^{\dagger}, \nonumber\\ \tilde{a}(\beta)
&=& e^{- i G} \tilde{a} e^{i G} = \cos \theta (\beta) \tilde{a} +
\sin \theta (\beta) a^{\dagger},
\end{eqnarray}
for fermions. Similar relations follow for $a^{\dagger} (\beta)$ and
$\tilde{a}^{\dagger} (\beta)$ for bosons and fermions.
The thermal state in Eq. (\ref{sq st}) is nothing but a
temperature-dependent vacuum state
\begin{equation}
a (\beta) \vert 0 (\beta) \rangle = \tilde{a} (\beta ) \vert 0
(\beta) \rangle = 0.
\end{equation}
Now the expectation value of an operator $A$ with respect to
thermal equilibrium is equivalent to that with respect to the
thermal state
\begin{equation}
\langle A \rangle_{\rm T} = \langle 0(\beta) \vert A \vert
0(\beta) \rangle.
\end{equation}
The expectation value of the number
operator leads to the boson distribution 
\begin{equation}
\langle 0 (\beta ) \vert a^{\dagger} a  \vert 0 (\beta) \rangle =
\sinh^2 \theta (\beta) = \frac{1}{e^{\beta \hbar \omega} - 1},
\end{equation}
and the fermion distribution
\begin{equation}
\langle 0 (\beta ) \vert a^{\dagger} a  \vert 0 (\beta) \rangle =
\sin^2 \theta (\beta) = \frac{1}{e^{\beta \hbar \omega} + 1}.
\end{equation}

\section{Time-Dependent Boson System}

We now consider a time-dependent boson system and find the
ingredients necessary for TFD extension. The most general time-dependent 
quadratic Hamiltonian for bosons takes the form
\begin{equation}
H(t) = \hbar \Bigl[ \omega_0 (t) a^{\dagger} a + \frac{1}{2} \omega_+(t) 
a^{\dagger 2} + \frac{1}{2} \omega_+^*a^2 \Bigr], \label{bos ham}
\end{equation}
where $a$ and $a^{\dagger}$ are the  Schr\"{o}dinger (time-independent)
annihilation and creation operators, and $\omega_0$ is real and $\omega_+$
real or complex. In particular, the oscillator with time-dependent 
mass and frequency, 
\begin{equation}
H (t) = \frac{p^2}{2m(t)} + \frac{m(t)}{2} \omega^2 (t)q^2,
\label{osc}
\end{equation}
belongs to the Hamiltonian, Eq. (\ref{bos ham}). Note that $p$ and
$q$ in Eq. (\ref{osc}) are also the Schr\"{o}dinger operators,
but the Hamiltonian depends explicitly on time through $m$ and
$\omega$. An instantaneous energy eigenstate $\vert e_n, t
\rangle$ of the Hamiltonian, Eq. (\ref{bos ham}), is definitely not an exact
quantum state of the Schr\"{o}dinger equation. Moreover, the
density operator is not given by $e^{- \beta H(t)}$. So the set $\{\vert
e_n, t \rangle \}$ is not a good basis to study the nonequilibrium
evolution of the thermal system.

We explain a physical motivation for replacing the Heisenberg
equation in the ingredient $(ii)$ by the LvN
equation. In the Heisenberg picture, it is the Heisenberg equation
that determines the operators evolving in time. In
fact, in terms of the evolution operator $U$ of the
Schr\"{o}dinger equation, each Schr\"{o}dinger
operator $A$ leads to the Heisenberg operator $A_{\rm H} =
U^{\dagger} A U$. For a time-dependent system, these operators
carry all the quantum information of the system.
As states do not evolve in time, the initial equilibrium
with the density operator $\rho$ leads to the thermal expectation
value of $A$ given by
\begin{equation}
\langle A \rangle_{\rm T} = {\rm Tr} [\rho A_{\rm H}(t)] = {\rm
Tr} [ U \rho U^{\dagger} A].
\end{equation}
Therefore, to know thermal properties of the system, either the
Heisenberg operator $A_{\rm H}$ or the density operator $\rho (t)
= U \rho U^{\dagger}$ should be given in advance.

On the other hand, there exist the invariant operators
that satisfy the LvN 
equation \cite{lewis}. A complete set of invariant
operators provide another picture for a time-dependent system and
such a set of invariant operators are known for time-dependent
oscillators in terms of classical solutions \cite{lin inv,kim2}. In this
sense the LvN equation may replace the Heisenberg
equation in TFD. Further, the linearity of the LvN 
equation allows the construction of the density operator from an invariant
operator \cite{kim1}. In fact, we may find the time-dependent 
annihilation operator, an invariant operator, of the form
\begin{equation}
a(t) = f^{(-)} (t) a + f^{(+)} (t) a^{\dagger} \label{bos an}
\end{equation}
and its Hermitian conjugate $a^{\dagger}(t)$, and impose the LvN equations
\begin{eqnarray}
i \hbar \frac{\partial a(t)}{\partial t} + [a (t), H(t)]_- = 0, \nonumber\\
i \hbar \frac{\partial a^{\dagger}(t)}{\partial t} + [a^{\dagger} (t), 
H(t)]_- = 0. \label{lvn eq}
\end{eqnarray}
The time-dependent creation operator $a^{\dagger} (t)$ is 
another invariant operator. 
The pair $\{ a(t), a^{\dagger} (t) \}$ form a complete set. Therefore
any invariant operator can be constructed out of them.

The LvN equation (\ref{lvn eq}) leads to the vector equation
\begin{equation}
i \frac{\partial V}{\partial t} + M_V V = 0, \label{bos vec}
\end{equation}
where
\begin{eqnarray}
 V (t) = \left( \begin{array}{c} f^{(-)} \\
f^{(+)} \end{array} \right),
\end{eqnarray}
and 
\begin{equation}
M_V (t) = \omega_0 \sigma_3 - \frac{1}{2} (\omega_+^* - \omega_+) 
\sigma_1 - \frac{i}{2} (\omega_+^* + \omega_+) \sigma_2.
\end{equation}
Here $\sigma$'s are the Pauli spin matrices. The solution to Eq. 
(\ref{bos vec}) provides the time-dependent annihilation and
creation operators. Further, the equal-time commutator
\begin{equation}
[a (t), a^{\dagger} (t)]_- = 1
\end{equation}
can hold by choosing the initial data 
$V^{\dagger} \sigma_3 V = 1$ since Eq. (\ref{bos vec}) leads to the relation
\begin{equation}
\frac{\partial}{\partial t} (V^{\dagger} \sigma_3 V ) = 0.
\end{equation}
For the time-dependent oscillator, Eq. (\ref{osc}), the following form of the
time-dependent annihilation operator is known \cite{lin inv,kim2}
\begin{eqnarray}
a (t) = \frac{i}{\sqrt{\hbar}} [v^*(t) p - m(t) \dot{v}^* q].
\end{eqnarray}
The time-dependent creation operator $a^{\dagger} (t)$ 
is the Hermitian conjugate of $a(t)$. 
In fact, these operators are invariant operators satisfying
the LvN equation (\ref{lvn eq}) for the Hamiltonian, Eq. (\ref{osc}),
when $v$ is a complex solution to the classical equation of motion
\begin{equation}
\ddot{v} (t) + \frac{\dot{m}(t)}{m(t)} v (t) + \omega^2(t) v (t) =
0, \label{cl eq}
\end{equation}
and satisfies the Wronskian condition
\begin{equation}
m ( \dot{v}^* v - \dot{v} v^*) = i. \label{wr con}
\end{equation}

The number states of the number operator
$N(t) = a^{\dagger} (t) a (t)$, another invariant operator, defined as,
\begin{equation}
N (t) \vert n, t \rangle = n \vert n, t \rangle,
\end{equation}
are exact quantum states of the Schr\"{o}dinger equation and
constitute the Hilbert space \cite{lewis,kim1,lin inv,kim2}. The state
$\vert 0, t \rangle$ is the time-dependent vacuum that is
annihilated by $a(t)$, and the number states are obtained by
applying the creation operators on it
\begin{equation}
\vert n, t \rangle = \frac{a^{\dagger n}(t)}{\sqrt{n!}} \vert 0, t
\rangle.
\end{equation}
It follows that the density operator may be given by
\begin{eqnarray}
{\rho} (t) = \frac{1}{Z} e^{- \beta \hbar \omega
{a}^{\dagger}(t) {a}(t)}, \label{den1}
\end{eqnarray}
where $\beta$ and $\omega$ are constants. In the case of the oscillator, 
Eq. (\ref{osc}), the position and momentum operators have 
the oscillator representation
\begin{eqnarray}
q &=& \sqrt{\hbar} [v (t) a (t) + v^* (t) a^{\dagger}
(t)],\nonumber\\ p &=& \sqrt{\hbar} m(t) [\dot{v} (t) a (t) +
\dot{v}^* (t) a^{\dagger} (t)]. \label{pos rep}
\end{eqnarray}
The merit of using the LvN equation is that we
can keep all the steps of the time-independent boson case in
finding the Hilbert space, the density operator and so on.

\section{TFD for Time-Dependent Boson System}

To extend TFD to the time-dependent boson system, Eq. (\ref{bos ham}), 
we use the tilde conjugation rule $(cA)^{\tilde{}} = c^* \tilde{A}$
and introduce a fictitious boson system with the Hamiltonian
\begin{equation}
\tilde{H}(t) = \hbar \Bigl[ \omega_0 (t) \tilde{a}^{\dagger} \tilde{a} 
+ \frac{1}{2} \omega_+^*(t) 
\tilde{a}^{\dagger 2} + \frac{1}{2} \omega_+
\tilde{a}^2 \Bigr]. \label{til osc}
\end{equation}
As for the time-dependent boson case, we introduce the
time-dependent annihilation operator for the
fictitious boson
\begin{equation}
\tilde{a} (t) = f^{(-)*} (t) \tilde{a} + f^{(+)*} (t) \tilde{a}^{\dagger},
\label{an ti}
\end{equation}
where $f^{(\pm)}$ satisfy Eq. (\ref{bos vec}). Then  $\tilde{a} (t)$ and
its Hermitian conjugate $\tilde{a}^{\dagger} (t)$ satisfy 
the LvN equations
\begin{eqnarray}
i \hbar \frac{\partial \tilde{a}(t)}{\partial t} + [\tilde{a} (t), - 
\tilde{H}(t)]_- = 0, \nonumber\\
i \hbar \frac{\partial \tilde{a}^{\dagger}(t)}{\partial t} + 
[\tilde{a}^{\dagger} (t), -
\tilde{H}(t)]_- = 0. \label{lvn eq til}
\end{eqnarray}
The equal-time commutator also holds
\begin{equation}
[ \tilde{a} (t), \tilde{a}^{\dagger} (t) ]_- = 1.
\end{equation}
The number states
\begin{equation}
\tilde{N} (t) \vert \tilde{n}, t \rangle = \tilde{a}^{\dagger} (t)
\tilde{a} (t) \vert \tilde{n}, t \rangle = \tilde{n} \vert
\tilde{n}, t \rangle
\end{equation}
are the exact quantum states for the Hamiltonian, Eq. (\ref{til osc}).

The total Hamiltonian is now given by
\begin{equation}
\hat{H} (t) = {H} (t) - \tilde{H} (t), \label{tot}
\end{equation}
where the operators of the boson and their tilde operators
commute with each other
\begin{equation}
[ \tilde{a}, {a} ]_- = [ \tilde{a}, {a}^{\dagger} ]_- = [ \tilde{
a}^{\dagger}, {a} ]_- = [ \tilde{a}^{\dagger}, {a} ]_- = 0,
\end{equation}
and
\begin{equation}
[ \tilde{a} (t), {a} (t) ]_- = [ \tilde{a} (t), 
{a}^{\dagger} (t) ]_- = [ \tilde{
a}^{\dagger} (t), {a} (t) ]_- = [ \tilde{a}^{\dagger} (t), {a} (t)]_- 
= 0.
\end{equation}
In fact, $a(t)$, $a^{\dagger} (t)$, $\tilde{a} (t)$ and $\tilde{a}^{\dagger} (t)$
are the invariant operators satisfying the LvN equations for the total
Hamiltonian, Eq. (\ref{tot}).
The Hilbert space of the total system consists of
\begin{eqnarray}
\vert n, \tilde{m}, t \rangle = \vert n, t \rangle \otimes \vert
\tilde{m}, t \rangle = \frac{{a}^{\dagger n}(t)}{\sqrt{n!}}
\frac{\tilde{a}^{\dagger m}(t)}{\sqrt{m!}} \vert
0, \tilde{0}, t \rangle.
\end{eqnarray}
The density operators may be defined by
\begin{eqnarray}
{\rho} (t) = \frac{1}{Z} e^{- \beta \hbar \omega
{a}^{\dagger}(t) {a}(t)}, \\ \tilde{\rho} (t) =
\frac{1}{Z} e^{+ \beta \hbar \omega \tilde{a}^{\dagger}(t)
\tilde{a}(t)}, \label{den2}
\end{eqnarray}
which obviously satisfy the LvN equations. Here
$\beta$ and $\omega$ are constants that may be fixed by the
initial temperature and frequency. The density operator in the
extended Hilbert space is given by
\begin{equation}
\hat{\rho} (t) = {\rho} (t) \otimes \tilde{\rho} (t) =
\frac{1}{Z^2} e^{ - \beta \hbar \omega ({a}^{\dagger}(t) {a}(t)
- \tilde{a}^{\dagger}(t) \tilde{a}(t) )}
\end{equation}
The thermal expectation value of the operator $A$ of the system
now takes the form
\begin{eqnarray}
\langle {A} \rangle = {\rm Tr} [{\rho} (t) {A}] = \langle
0(\beta), t \vert {A} \vert 0(\beta), t \rangle,
\end{eqnarray}
where the thermal vacuum state is given by
\begin{eqnarray}
\vert 0(\beta), t \rangle &=& \frac{1}{Z^{1/2}} \sum_{n} e^{-
\beta \hbar \omega n/2} \frac{1}{n!} {a}^{\dagger n} (t)
\tilde{a}^{\dagger \tilde{n}}(t) \vert 0, \tilde{0}, t \rangle \nonumber\\
&=& \sqrt{1 - e^{- \beta \hbar \omega}} e^{- (\beta \hbar
\omega /2) {a}^{\dagger} (t) \tilde{a}^{\dagger} (t)} \vert 0, t
\rangle,
\end{eqnarray}
with $\vert 0, t \rangle = \vert 0, \tilde{0}, t \rangle$. 
The thermal state is an exact eigenstate of the Schr\"{o}dinger 
equation for the total system, Eq. (\ref{tot}).
The thermal state is also written as a time-dependent two-mode
squeezed state of the vacuum state
\begin{equation}
\vert 0(\beta), t \rangle = e^{-i G (t)} \vert 0, t \rangle,
\end{equation}
where
\begin{equation}
G (t) = - i \theta (\beta) [ \tilde{a} (t) a (t) - a^{\dagger} (t)
\tilde{a}^{\dagger} (t) ].
\end{equation}
Here $\theta (\beta)$ is the same parameter fixed by Eq. (\ref{cosh}).

As the density operator in Eq. (\ref{den1}) or (\ref{den2}) involves a
constant $\beta$, we may find the time- and temperature-dependent
annihilation and creation operators through the Bogoliubov
transformation
\begin{eqnarray}
a(\beta, t) &=& \cosh \theta (\beta) a (t) - \sinh \theta (\beta)
\tilde{a}^{\dagger} (t), \nonumber\\ \tilde{a}(\beta, t) &=& \cosh
\theta (\beta) \tilde{a} (t) - \sinh \theta (\beta) a^{\dagger}
(t), \label{tem bog}
\end{eqnarray}
and their inverse transformation
\begin{eqnarray}
a (t) &=& \cosh \theta (\beta) a (\beta, t) + \sinh \theta (\beta)
\tilde{a}^{\dagger} (\beta, t), \nonumber\\ \tilde{a} (t) &=&
\cosh \theta (\beta) \tilde{a} (\beta, t) + \sinh \theta (\beta)
a^{\dagger} (\beta, t). \label{inv bog}
\end{eqnarray}
We get similar equations for $a^{\dagger}(\beta, t)$, 
$\tilde{a}^{\dagger} (\beta, t)$,
$a^{\dagger} (t)$ and $\tilde{a}^{\dagger} (t)$ by using the Hermitian 
conjugate of these equations. As $\theta (\beta)$ is a constant, $a(\beta, t)$,
$\tilde{a}(\beta, t)$ are invariant operators.
Then the thermal state is the time- 
and temperature-dependent vacuum
\begin{equation}
a (\beta, t) \vert 0 (\beta), t \rangle = \tilde{a} (\beta, t)
\vert 0 (\beta), t \rangle = 0. \label{th con}
\end{equation}
The thermal state $\vert 0(\beta), t \rangle$, as an eigenstate of the
invariant operators $a(\beta, t)$ and $\tilde{a}(\beta, t)$, is an exact 
eigenstate of the total system.
At each moment, the boson still keeps the same boson
distribution since the expectation value of the time-dependent
number operator yields
\begin{equation}
\langle 0 (\beta ), t \vert a^{\dagger} (t) a (t)  \vert 0
(\beta), t \rangle = \sinh^ 2 \theta (\beta) = \frac{1}{e^{\beta
\hbar \omega} - 1}.
\end{equation}

Using TFD for time-dependent bosons, we are able to find
the thermal expectation values of operators.
In general, through the Bogoliubov transformations from $\{ a(t),
a^{\dagger} (t)\}$ to $\{ a(\beta, t), a^{\dagger} (\beta, t)\}$,
we find the formula
\begin{eqnarray}
\langle F(a(t), a^{\dagger} (t)) \rangle_{\rm T} &=& \langle 0
(\beta), t  \vert F(\cosh \theta (\beta) a (\beta, t) + \sinh
\theta (\beta) \tilde{a}^{\dagger} (\beta, t), \nonumber\\&& \cosh
\theta (\beta) a^{\dagger} (\beta, t) + \sinh \theta (\beta)
\tilde{a} (\beta, t)) \vert 0 (\beta), t \rangle.
\end{eqnarray}
This provides the basic rule for calculating matrix element 
of any operator in TFD. For instance, 
in the case of the oscillator, Eq. (\ref{osc}),
using the position representation in Eq. (\ref{pos rep})
\begin{eqnarray}
q &=& \sqrt{\hbar} \cosh \theta (\beta) [ v(t) a( \beta, t) + v^*
(t) a^{\dagger} (\beta, t)] \nonumber\\ &&+ \sqrt{\hbar} \sinh
\theta (\beta) [ v^*(t) \tilde{a} (\beta, t) + v(t)
\tilde{a}^{\dagger} (\beta, t)],
\end{eqnarray}
we obtain
\begin{eqnarray}
\langle 0 (\beta), t \vert q^{2n} \vert 0 (\beta), t \rangle &=&
\hbar^n \sum_{k = 0}^{n} {2n \choose 2k} \langle 0, t \vert
\cosh^{2k} \theta (\beta) [ v(t) a(\beta, t) + v^* (t) a^{\dagger}
(\beta, t)]^{2k} \nonumber\\ && \times \sinh^{2n - 2k} \theta
(\beta) [ v^*(t) \tilde{a}(\beta, t) + v(t) \tilde{a}^{\dagger}
(\beta, t)]^{2n-2k} \vert 0, t \rangle.
\end{eqnarray}
After normal ordering, we finally obtain the result
\begin{eqnarray}
\langle q^{2n} \rangle_{\rm T} &=& \langle 0 (\beta), t \vert
q^{2n} \vert 0 (\beta), t \rangle \nonumber\\ &=& \frac{(2n)!}{2^n
n!} [\hbar v^* (t) v(t)]^{n} ( 1 + 2 \sinh^2 \theta (\beta))^n.
\end{eqnarray}

We discuss the physical implication of TFD for a time-dependent
boson, when it evolves with a time-dependent interaction
from initial $\omega_i$'s at $t = t_i$ to final ones,
$\omega_f$'s at $t_f$. That is, all $\omega$'s change from $\omega_i$'s
to $\omega_f$'s. The solution to Eq. (\ref{bos vec}) is necessary in
finding the time-dependent annihilation operators in
Eqs. (\ref{bos an}) and (\ref{an ti}) and their Hermitian conjugates. 
However, from constants $\omega_f$'s, we may find a
Bogoliubov transformation of the form
\begin{eqnarray}
a_i &=& \mu a_f + \nu a_f^{\dagger}, \nonumber\\ a_i^{\dagger} &=&
\mu^* a^{\dagger}_f + \nu^* a_f,
\end{eqnarray}
where $\{a_i, a_i^{\dagger}\}$ for $\omega_i$'s and $\{a_f,
a_f^{\dagger}\}$ for $\omega_f$'s. Here $\mu$ and $\nu$, which
should be determined by the solution $f^{(\pm)}$ to Eq. (\ref{bos vec}),
carry all the information about the history of interaction and may
take the form
\begin{equation}
\mu = \mu (t_i, t_f; \omega_i, \omega_f), \quad \nu = \nu (t_i,
t_f; \omega_i, \omega_f),
\end{equation}
and satisfy
\begin{equation}
\mu^* \mu - \nu^*  \nu = 1.
\end{equation}
If the boson is initially in thermal equilibrium with the
inverse temperature $\beta$ and has the boson distribution
$\bar{n}_i = 1/(e^{\beta \hbar \omega} -1)$, then, according to
Secs. III and IV, the boson in the final state has a
different distribution
\begin{eqnarray}
\langle 0 (\beta), t_f \vert a_i^{\dagger} a_i \vert 0(\beta), t_f
\rangle = \nu^* \nu + \frac{1 + 2 \nu^* \nu}{e^{\beta \hbar
\omega} - 1}.
\end{eqnarray}
The first term is originated from the particle production from vacuum fluctuations
\cite{parker}, $\langle 0, t_f \vert a_i^{\dagger} a_i \vert 0,
t_f \rangle = \nu^* \nu$,  and the second term is a purely thermal
result, having an overall amplification factor $(1 +
2 \nu^* \nu)$ to the boson distribution. Thus the evolution of the 
time-dependent system leads to a distribution quite different from the boson 
distribution function.

\section{TFD for Time-Dependent Fermion System}

The time-dependent fermion system, quadratic in the annihilation 
and creation operators, has the Hamiltonian
\begin{equation}
H(t) = \hbar [ \omega_0 (t) (a^{\dagger} a - b^{\dagger} b) + \omega_+(t) 
a^{\dagger} b^{\dagger} - \omega_+^*(t) ab + \omega_- (t) a b^{\dagger} -
\omega_-^*(t) a^{\dagger} b], \label{fer ham}
\end{equation}
where $a, a^{\dagger}$ for the particles 
and $b, b^{\dagger}$ for the antiparticles are all the 
Schr\"{o}dinger (time-independent) operators, and $\omega_0$ is real and 
$\omega_{\pm}$ are real or complex. They satisfy the anticommutators
\begin{eqnarray}
[ a, a^{\dagger} ]_+ = [ b, b^{\dagger} ]_+ = 1, \quad [ a, b ]_+ = [ a, b^{\dagger} ]_+ = [ a^{\dagger}, b^{\dagger} ]_+ = 0.
\end{eqnarray}
The Hamiltonian, Eq. (\ref{fer ham}), is a Hermitian operator and its
time-dependency comes only from the parameters $\omega$'s.

As for the boson case, we may find a pair of time-dependent invariant annihilation
operators \cite{kim-khanna}
\begin{eqnarray}
a (t) &=& f_a^{(-)} (t) a + f_a^{(+)} (t) a^{\dagger} 
+ g_a^{(-)} (t) b + g_a^{(+)} (t) b^{\dagger}, \nonumber\\
b (t) &=& f_b^{(-)} (t) a + f_b^{(+)} (t) a^{\dagger} 
+ g_b^{(-)} (t) b + g_b^{(+)} (t) b^{\dagger}, \label{fer op}
\end{eqnarray}
and the invariant creation operators $a^{\dagger} (t)$ and 
$b^{\dagger} (t)$ are the Hermitian conjugates of $a(t)$ and $b(t)$, respectively.
The LvN equations for the operators in Eq.
(\ref{fer op}) with the Hamiltonian, Eq. (\ref{fer ham}), lead to the following 
vector equations \cite{kim-khanna}
\begin{eqnarray}
i \frac{\partial W}{\partial t} + \omega_0 \sigma_1 W + M_W Z = 0, \nonumber\\
i \frac{\partial Z}{\partial t} + \omega_0 \sigma_1 Z + M_Z W = 0, \label{vec eq}
\end{eqnarray}
where
\begin{eqnarray}
 W (t) &=& \frac{1}{\sqrt{2}} \left( \begin{array}{c} f^{(-)} + f^{(+)} \\
f^{(-)} - f^{(+)} \end{array} \right), \nonumber\\
Z (t) &=& \frac{1}{\sqrt{2}} \left( \begin{array}{c} g^{(-)} + g^{(+)} \\
g^{(-)} - g^{(+)} \end{array} \right),
\end{eqnarray}
and
\begin{eqnarray}
M_W (t) &=& \frac{1}{2} (\omega_-^* - \omega_-) I 
- \frac{1}{2} (\omega_-^* + \omega_-) 
\sigma_1 - \frac{i}{2} (\omega_+^* + \omega_+) \sigma_2 - \frac{1}{2} (\omega_+^*
- \omega_+) \sigma_3, \nonumber\\ 
M_Z (t) &=& - \frac{1}{2} (\omega_-^* - \omega_-) I 
- \frac{1}{2} (\omega_-^* + \omega_-) 
\sigma_1 + \frac{i}{2} (\omega_+^* + \omega_+) \sigma_2 + \frac{1}{2} (\omega_+^*
- \omega_+) \sigma_3,
\end{eqnarray}
where $I$ is the identity matrix. The 
equal-time anticommutators
\begin{eqnarray}
[ a (t), a^{\dagger} (t) ]_+ = [ b (t), b^{\dagger} (t) ]_+ = 1, 
\quad [ a (t), b (t)]_+ = [ a(t), b^{\dagger}(t) ]_+ = [ a^{\dagger} (t), 
b^{\dagger} (t) ]_+ = 0,
\end{eqnarray}
are guaranteed by the relations
\begin{eqnarray}
\frac{\partial}{\partial t} (W^{\dagger} W + Z^{\dagger} Z) = 0, \nonumber\\
\frac{\partial}{\partial t} (W_a^T \sigma_3 W_b + Z_a^T \sigma_3 Z_b) = 0, 
\nonumber\\ \frac{\partial}{\partial t} (W_a^{\dagger} W_b 
+ Z_a^{\dagger} Z_b) = 0.
\end{eqnarray}   
Each set of $\{W_a, Z_a \}$ and $\{W_b, Z_b \}$ gives rise to 
$a(t)$ and $b(t)$, and their 
Hermitian conjugates. The number operators $N_a (t) = a^{\dagger} (t) a(t)$ 
and $N_b (t) 
= b^{\dagger} (t) b(t)$, which are also invariant operators, 
span the state vector space 
of the fermions
\begin{equation}
\vert 0, t \rangle, \quad a^{\dagger} (t) \vert 0, t \rangle, 
\quad b^{\dagger} (t) \vert 0, t \rangle, \quad a^{\dagger} (t) b^{\dagger} (t) 
\vert 0, t \rangle.
\end{equation}
Here the vacuum state, $\vert 0, t \rangle = \vert 0, \tilde{0}, t \rangle$, 
implies no particles and antiparticles. 
The density operator may take the form
\begin{equation}
\rho (t) = \frac{1}{Z} e^{- \beta \hbar \omega (a^{\dagger} (t) 
a(t) - b^{\dagger}(t) b (t))},
\end{equation}
where $\beta$ and $\omega$ are again constants.

To construct the TFD for fermions, we introduce the fictitious fermion Hamiltonian
\begin{equation}
\tilde{H} (t) = \hbar [ \omega_0 (t) (\tilde{a}^{\dagger} \tilde{a} 
- \tilde{b}^{\dagger} \tilde{b}) + \omega_+^*(t) 
\tilde{a}^{\dagger} \tilde{b}^{\dagger} - \omega_+(t) \tilde{a}
\tilde{b} + \omega_-^* (t) \tilde{a} \tilde{b}^{\dagger} -
\omega_-(t) \tilde{a}^{\dagger} \tilde{b}]. \label{fer ham til}
\end{equation}
Then there are the time-dependent annihilation operators, invariant operators,
\begin{eqnarray}
\tilde{a} (t) &=& f_a^{(-)*} (t) \tilde{a} + f_a^{(+)*} (t) \tilde{a}^{\dagger} 
+ g_a^{(-)*} (t) \tilde{b} + g_a^{(+)*} (t) \tilde{b}^{\dagger}, \nonumber\\
\tilde{b} (t) &=& f_b^{(-)*} (t) \tilde{a} + f_b^{(+)*} (t) \tilde{a}^{\dagger} 
+ g_b^{(-)*} (t) \tilde{b} + g_b^{(+)*} (t) \tilde{b}^{\dagger}. \label{fer op til}
\end{eqnarray}
These operators satisfy the LvN equations (\ref{lvn eq til})
when $f$'s and $g$'s satisfy 
the vector equations (\ref{vec eq}). We are then equipped with the time-dependent 
annihilation and creation operators for the total system 
\begin{equation}
\hat{H} (t) = H(t) - \tilde{H} (t).
\end{equation}
The time-dependent vacuum state of the total system is annihilated by all 
the annihilation operators of the fermions and the fictitious fermions: 
\begin{equation}
a(t) \vert 0, \tilde{0}, t \rangle = b(t) \vert 0, \tilde{0}, t \rangle
= \tilde{a}(t) \vert 0, \tilde{0}, t \rangle 
= \tilde{b}(t) \vert 0, \tilde{0}, t \rangle = 0.
\end{equation}
Here the vacuum state, $\vert 0, \tilde{0}, t \rangle = \vert 0, 0, \tilde{0},
\tilde{0}, t \rangle$, implies no particles or antiparticles and their 
counterparts. However, we will continue to use the abbreviation $\vert 0, 
\tilde{0}, t \rangle$ for this state.

The thermal state of TFD is the two-mode squeezed state of the time-dependent 
vacuum state
\begin{equation}
\vert 0 (\beta), t \rangle = e^{- i G_F (t)} \vert 0, t \rangle,
\end{equation}
where $\vert 0, t \rangle = \vert 0, \tilde{0}, t \rangle$, and
\begin{equation}
G_F (t) = -i \theta (\beta) (\tilde{a} a - a^{\dagger} \tilde{a}^{\dagger} 
+ \tilde{b} b - b^{\dagger} \tilde{b}^{\dagger}).
\end{equation}
Here $\theta (\beta)$ is the same parameter fixed by Eq. (\ref{cos}).
The two-mode squeeze operator introduces the time- 
and temperature-dependent annihilation 
and creation operators for the particles through the Bogoliubov transformation
\begin{eqnarray}
a(\beta, t) &=& \cos \theta (\beta) a (t) - \sin \theta (\beta)
\tilde{a}^{\dagger} (t), \nonumber\\ \tilde{a}(\beta, t) &=& \cos
\theta (\beta) \tilde{a} (t) + \sin \theta (\beta) a^{\dagger}
(t), \label{fer tem}
\end{eqnarray}
and for the antiparticles
\begin{eqnarray}
b(\beta, t) &=& \cos \theta (\beta) b (t) - \sin \theta (\beta)
\tilde{b}^{\dagger} (t), \nonumber\\ \tilde{b}(\beta, t) &=& \cos
\theta (\beta) \tilde{b} (t) + \sin \theta (\beta) b^{\dagger}
(t). \label{fer tem2}
\end{eqnarray}
We get similar equations for $a^{\dagger}(\beta, t), \tilde{a}^{\dagger}(\beta, t),
b^{\dagger}(\beta, t)$ and $\tilde{b}^{\dagger}(\beta, t)$ by using Hermitian 
conjugate on these equations. Their inverse transformations are
\begin{eqnarray}
a (t) &=& \cos \theta (\beta) a (\beta, t) + \sin \theta (\beta)
\tilde{a}^{\dagger} (\beta, t), \nonumber\\ \tilde{a} (t) &=&
\cos \theta (\beta) \tilde{a} (\beta, t) - \sin \theta (\beta)
a^{\dagger} (\beta, t)
\end{eqnarray}
and
\begin{eqnarray}
b (t) &=& \cos \theta (\beta) b (\beta, t) + \sin \theta (\beta)
\tilde{b}^{\dagger} (\beta, t), \nonumber\\ \tilde{b} (t) &=&
\cos \theta (\beta) \tilde{b} (\beta, t) - \sin \theta (\beta)
b^{\dagger} (\beta, t).
\end{eqnarray}
Further, the thermal state is annihilated by
\begin{equation}
a(\beta, t ) \vert 0(\beta), t \rangle = \tilde{a} (\beta, t ) 
\vert 0(\beta), t \rangle
= b(\beta, t ) \vert 0(\beta), t \rangle 
= \tilde{b}(\beta, t ) \vert 0(\beta), t \rangle.
\end{equation}
Now the expectation value of the particle number operator
\begin{equation}
\langle 0(\beta), t \vert a^{\dagger} (t) a(t) \vert 0(\beta), t \rangle 
= \sin^2 \theta = \frac{1}{e^{\beta \hbar \omega} + 1}.
\end{equation}
The thermal expectation value of operators, for instance, of the particles is given
by the formula
\begin{eqnarray}
\langle F(a(t), a^{\dagger} (t)) \rangle_{\rm T} &=& \langle 0
(\beta), t  \vert F(\cos \theta (\beta) a (\beta, t) + \sin
\theta (\beta) \tilde{a}^{\dagger} (\beta, t), \nonumber\\&& \cos
\theta (\beta) a^{\dagger} (\beta, t) + \sin \theta (\beta)
\tilde{a} (\beta, t)) \vert 0 (\beta), t \rangle.
\end{eqnarray}

\section{Conclusion}

To summarize, in this paper we have used the time-dependent
annihilation and creation operators of the LvN
equation to complete TFD for time-dependent boson and fermion systems. 
The first ingredient, $(i)$ the tilde conjugation rule, has been
accomplished by introducing the fictitious boson or fermion operators and
appropriately constructing the extended Hilbert space. The second
ingredient, $(ii)$ the Heisenberg equation, is replaced
by the LvN equation. The annihilation
and creation operators from the LvN equation
have led to exactly the same procedures as for TFD of
time-independent boson and fermion systems. The third ingredient, $(iii)$ the
thermal state condition, is guaranteed by the time- and
temperature-dependent Bogoliubov transformations, Eqs. (\ref{tem bog}), 
(\ref{fer tem}) and (\ref{fer tem2}). The thermal state for time-dependent 
bosons is the time- and temperature-dependent vacuum as 
in Eq. (\ref{th con}), which can be written as
\begin{equation}
a (t) \vert 0(\beta), t \rangle = \tanh \theta (\beta)
\tilde{a}^{\dagger} (t) \vert 0(\beta), t \rangle, \quad \tilde{a}
(t) \vert 0(\beta), t \rangle = \tanh \theta (\beta) a^{\dagger}
(t) \vert 0(\beta), t \rangle. \label{ther con}
\end{equation}
The thermal state condition for time-dependent fermions is given by
\begin{eqnarray}
a (t) \vert 0(\beta), t \rangle &=& \tan \theta (\beta)
\tilde{a}^{\dagger} (t) \vert 0(\beta), t \rangle, \quad \tilde{a}
(t) \vert 0(\beta), t \rangle = - \tan \theta (\beta) a^{\dagger}
(t) \vert 0(\beta), t \rangle, \nonumber\\
b (t) \vert 0(\beta), t \rangle &=& \tan \theta (\beta)
\tilde{b}^{\dagger} (t) \vert 0(\beta), t \rangle, \quad \tilde{b}
(t) \vert 0(\beta), t \rangle = - \tan \theta (\beta) b^{\dagger}
(t) \vert 0(\beta), t \rangle. \label{ther con2}
\end{eqnarray}
Note that the thermal state conditions, Eqs. (\ref{ther con}) and (\ref{ther
con2}), still have the same form as for the time-independent boson and fermion 
systems. We may thus conclude that the LvN equation
provides a direct generalization of TFD to nonequilibrium systems such as
time-dependent boson and fermion systems. 
Finally it should be stressed that
the replacement of the Heisenberg equation by the LvN 
equation leads us from the Heisenberg picture to the Schr\"{o}dinger picture. 
Such a procedure yields interesting and useful results for quantum mechanical 
time-dependent systems. Extension to quantum field theory 
may be possible and this would bring
the open systems to a treatment similar to the case of systems in equilibrium. 
That would provide a true extension of TFD, a la Takahashi and Umezawa that 
has proved so very successful for treating equilibrium problems, to systems
out of equilibrium. Hopefully this will provide a useful perspective on such 
a class of problems. This topic is under active consideration.

\acknowledgements

We would like to thank M. Revzen for useful discussions.
S.P.K. also would like to express his appreciation for the warm hospitality
of the Theoretical 
Physics Institute, University of Alberta. The work
of S.P.K. was supported by the Korea Research Foundation under
Grant No. KRF-2002-041-C00053.

\end{document}